\documentclass[aps,prb,twocolumn,showpacs,groupedaddress]{revtex4-1}
\usepackage{latexsym}
\usepackage[english]{babel}
\usepackage{graphicx}
\usepackage{amsfonts}
\usepackage{amsmath}
\usepackage{amssymb}
\usepackage{slashed}
\usepackage{verbatim}
\usepackage{feynmf}
\usepackage{mathrsfs}
\usepackage{amsmath,amssymb}
\usepackage{epsfig}
\usepackage{relsize}
\usepackage{graphicx}
\input epsf
\usepackage{bbm}
\usepackage{color}

\begin{document}

\title{Spin- and band-ferromagnetism in trilayer graphene}

\author{Ralph van Gelderen}
\email{R.vangelderen1@uu.nl}

\affiliation{Institute for Theoretical Physics, Utrecht
University, Leuvenlaan 4, 3584 CE Utrecht, The Netherlands}

\author{Lih-King Lim}

\affiliation{Institute for Theoretical Physics, Utrecht
University, Leuvenlaan 4, 3584 CE Utrecht, The Netherlands}

\author{C. Morais Smith}

\affiliation{Institute for Theoretical Physics, Utrecht
University, Leuvenlaan 4, 3584 CE Utrecht, The Netherlands}

\date{\today}


\begin{abstract}
We study the ground state properties of an ABA-stacked trilayer graphene. The low energy band structure can be described by a combination of both a linear and a quadratic particle-hole symmetric dispersions, reminiscent of monolayer- and bilayer-graphene, respectively. The multi-band structure offers more channels for instability towards ferromagnetism when the Coulomb interaction is taken into account. Indeed, if one associates a pseudo-spin $1/2$ degree of freedom to the bands (parabolic/linear), it is possible to realize also a band-ferromagnetic state, where there is a shift in the energy bands, since they fill up differently.
By using a variational procedure, we compute the exchange energies for all possible variational ground states and identify the parameter space for the occurrence of spin- and band-ferromagnetic instabilities as a function of doping and interaction strength.
\end{abstract}
%\vskip2pc

\maketitle

\section{Introduction}

The successful isolation of a one atom thick carbon layer, graphene, has attracted enormous interest in the field of condensed matter.\cite{Novo04,Castro09}
One intriguing aspect of the problem is that upon coupling a finite number of graphene layers, novel and unexpected properties emerge.
%However, the studies of graphene are not restricted to a single layer but also to a stack of graphene sheets.
Compared to the strong $sp^2$ bonding between carbon atoms within the graphene sheet, the weak van der Waals force between the layers allows for the formation of different hybridized $N$-layered configurations. The resulting system is then different from both its 2D (graphene) and its 3D (graphite) counterparts, and depends strongly on the number of layers and on how the stacking is realized. %remains effectively \emph{two-dimensional} as long as $N$ remains small and the additional degree of freedom stems from the way in which different stacking can be realized.
The investigation of multi-layer graphene may open new avenues in the understanding of graphene's electronic properties and in the field of device engineering.\cite{McCa10a,McDo08a}

Many of the unique electronic properties of monolayer graphene, as opposed to the more conventional GaAs 2D electron gas, originate from the geometry of the honeycomb lattice. These include the peculiar gapless Dirac-cone dispersion,\cite{Castro09} the unconventional integer quantum Hall effect,\cite{Geim05} and Klein tunneling,\cite{NoGe06} to name a few. On the other hand, multi-layer graphene exhibits different but equally interesting features. While the particle-hole symmetry is generally preserved in the band structure obtained from the minimal tight-binding description, the number of conical points and the low-energy dispersion both depend sensitively on the stacking configuration of the $N$-layered structure.  For example, in the so-called Bernal stacking of a bilayer graphene, the
conduction and valence bands touch at the same two points in the Brillouin zone as they do in monolayer graphene, but disperse quadratically instead of linearly. This feature has attracted much interest because it allows for strong electron correlations to take place.\cite{Zhang10,Vafek10} Very recently, broken-symmetry states have
been observed due to interaction effects in suspended bilayer graphene.\cite{Feldman09,Weitz10,Freitag11} Although a complete characterization of their properties is still lacking, there are some interesting theoretical proposals
for the observed states:  a many-body excitonic state\cite{Zhang10,Vafek10} or an anomalous spin-Hall state with time-reversal symmetry.\cite{NandPRL10,NandPRB10} For another example, the relative twist angle in a bilayer graphene can lead to a highly complex Moir\'{e} band structure, which requires a description beyond the standard Bloch's band picture. In fact, at a particular twisting angle, the van Hove singularity of the usual graphene band structure can become observable at a relatively low energy of a few meV.\cite{Eva11}
 Since high quality samples of $N$-layered graphene are now becoming accessible experimentally, their anticipated new properties are just about to be unraveled.
 %The interesting properties of bilayer graphene hold promises that a similarly complex behavior may arise in trilayer samples.

\begin{figure}[t!]
\includegraphics[width=.25\textwidth]{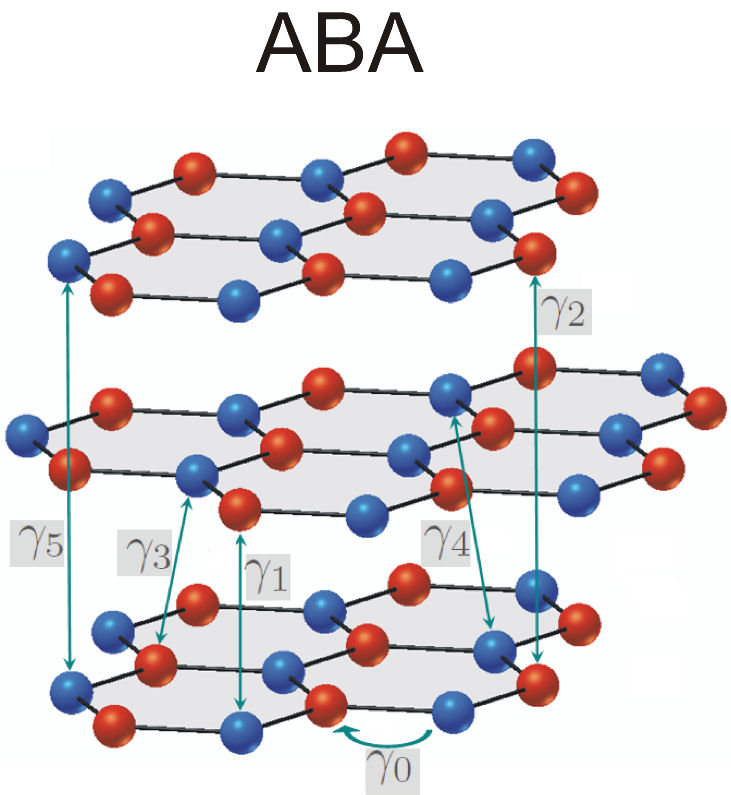}
\caption{(Color online) ABA stacked trilayer graphene with the various hopping parameters.}
\label{fig1}
\end{figure}

In trilayer graphene, the transport properties are also different, depending on the stacking order:
at the Dirac point, the ABA-stacked trilayer (Fig.~\ref{fig1}) is a semimetal, whereas the ABC one is a semiconductor, with an
intrinsic band gap.\cite{Bao11} The electronic band structure in ABC-stacked trilayer graphene was
determined using an effective mass approximation\cite{Koshino2} and using an ab-initio density functional theory.\cite{FanZhang10} On the other hand, for ABA-stacked the band structure was
calculated in the presence of external gates using a self-consistent Hartree approximation.\cite{Koshino1} In the absence of a gate, the low-energy spectrum consists of superimposed linear and quadratic bands, which touch at ${\bf k} = 0$.
In the presence of a magnetic field, the plateau structure in the Hall
conductivity is also determined by the stacking order. Very recently, the integer quantum Hall effect was experimentally observed in an ABC-stacked sample.\cite{LiyuanZhang11,Kumar11} It was shown that the effect is similar to the one observed in monolayer graphene,\cite{Geim05} except for the first plateau at filling factor $\nu = 2$, which was not observed in the trilayer sample. Indeed, this plateau is governed by the chirality of the quasi-particles, which is 1, 2, and 3 for monolayer, bilayer, and trilayer graphene, respectively. The corresponding Berry phases are thus $\pi$, $2 \pi$, and $3 \pi$, respectively.
With regard to the ABA stacking, the problem of low mobility has been recently overcome, by growing the sample on a high-quality hexagonal boron nitride substrate, which reduces the carrier scattering.\cite{Taycha11} The peculiar crossing of the Landau levels due to the massive and massless  sub-bands has allowed for a direct determination of the Slonczweski-Weiss-McClure model parameters used to describe the electronic structure of the material.\cite{SW58,McC57}

\begin{figure}[h!]
\includegraphics[width=.4\textwidth]{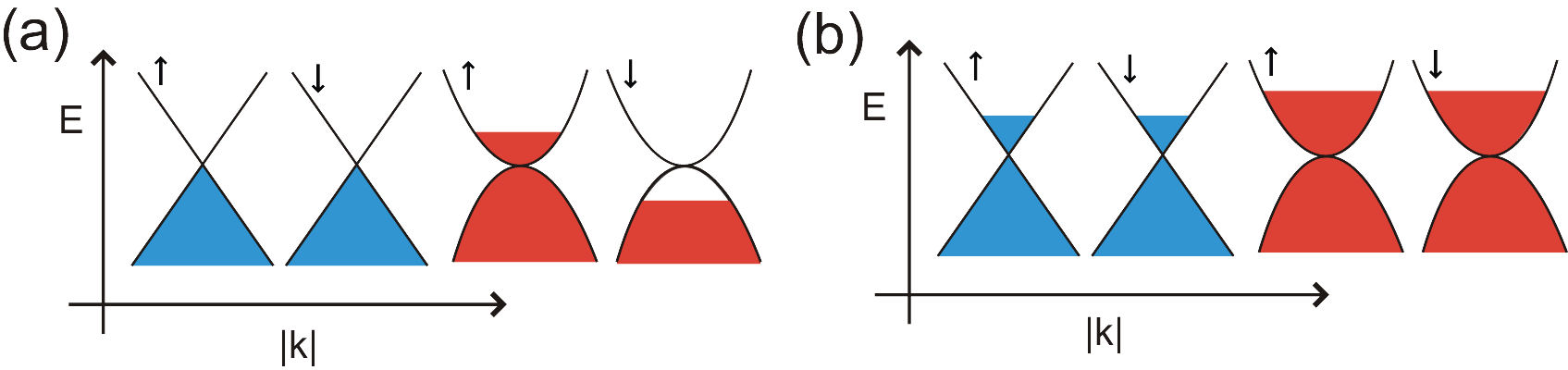}
\caption{(Color online) Sketch of (a) the spin-ferromagnetic state in an undoped trilayer and (b) the band-ferromagnetic state in a doped trilayer.}
\label{fig2}
\end{figure}

We focus here on the ground state properties of trilayer graphene in the ABA-stacking configuration in the presence of interactions. The ground state of $N$-layer undoped graphene is usually assumed to be the state in which the energy bands are filled up to the Dirac point. However, the energy bands are spin degenerate and the formation of pockets of opposite sign in the two spin degenerate bands leads to a gain in exchange energy. This gain in exchange energy is accompanied by a cost in kinetic energy. In monolayer graphene, the cost in kinetic energy is large enough to prevent any ferromagnetic instabilities;\cite{PeCaNe05a} only if the interaction would be tuned to unphysical values one would observe the spontaneous generation of spin up and spin down pockets. In bilayer graphene the situation is different. The leading order term in the exchange energy is one order lower in the pocket size than the kinetic energy is. Therefore, the exchange interaction dominates and pockets will form with a size, in $k$-space, of order $Q \approx 0.05 t_\perp$, where $t_\perp$ is the interlayer hopping energy in dimensionless units and $Q$ is measured in units of some cut off.\cite{NiCaNe06a} Hence, bilayer graphene has a small ferromagnetic instability.
The coexistence of a parabolic and a linear bands in ABA-trilayer graphene opens the way to investigate, next to ordinary ferromagnetic instabilities (Fig.~\ref{fig2}a), also the 'band ferromagnetism' phenomenon. With band ferromagnetism we mean that the two bands (linear and parabolic) become shifted with respect to each other (the crossing point of the linear and parabolic conduction and valence bands no longer overlap), or alternatively, that the bands fill up to different Fermi energies (see Fig.~\ref{fig2}b). In the following, we will generalize the approach used in Refs.~\onlinecite{PeCaNe05a} and \onlinecite{NiCaNe06a} to investigate ferromagnetic instabilities in trilayer graphene. We will show that spin- and band-ferromagnetism may occur both separately and simultaneously. %(Fig.~\ref{fig2}c).
The paper is organized as follows: In section II we introduce the model that we use in section III to compute (band) ferromagnetic instabilities for both undoped and doped trilayer graphene. Our conclusions are presented in section IV.

%\begin{figure}
%\includegraphics[width=.48\textwidth]{dopedpockets1.eps}
%\includegraphics[width=.48\textwidth]{dopedpockets2.eps}
%\caption{Band ferromagnetism: The linear and parabolic bands can shift with respect to each other, such that electron pockets (red) are formed in one %band and hole pockets (blue) in the other one. One has to calculate the total energy to determine which configuration is favorable.}
%\label{fig1}
%\end{figure}

\section{The Model}

In this paper, we use a tight-binding approximation to model trilayer graphene and perform an expansion around the $K$ point. The low-energy Hamiltonian around the $K$ point is given by $$H=\sum \Psi_{\mathbf{p},\sigma}^\dagger \mathcal{H}(
\mathbf{p}) \Psi_{\mathbf{p},\sigma},$$ where $\Psi_{
\mathbf{p},
\sigma}^\dagger = ( a_{1,\mathbf{p},\sigma}^\dagger,b_{1,\mathbf{p},\sigma}^\dagger,a_{2,\mathbf{p},\sigma}^\dagger,b_{2,
\mathbf{p},\sigma}^\dagger,a_{3,\mathbf{p},\sigma}^\dagger,b_{3,\mathbf{p},\sigma}^\dagger)$,
\begin{widetext}
\begin{align}
\label{ham1} \mathcal{H}(
\mathbf{p})&=\left( \begin{array}{cccccc} 0 & v_F p e^{i \phi(\mathbf{p})} &0& -t_\perp &0&0 \\ v_F p e^{-i \phi(\mathbf{p})} & 0&0&0&0&0 \\ 0&0&0& v_F p e^{i \phi(\mathbf{p})} &0&0 \\ -t_\perp &0& v_F p e^{-i \phi(\mathbf{p})} &0&-t_\perp &0 \\ 0&0&0& -t_\perp &0& v_F p e^{i \phi(\mathbf{p})} \\ 0&0&0&0& v_F p e^{-i \phi(\mathbf{p})} &0 \end{array} \right), \end{align}
\end{widetext}
and the sum is over all relevant quantum numbers. Here, $a^\dagger_{i,\mathbf{p},\sigma}$ ($b^\dagger_{i,\mathbf{p},\sigma}$) creates a particle with momentum $
\mathbf{p}$ and spin $
\sigma$ at the $A$ ($B$) sublattice in the $i$-th layer ($i=1,2,3$), $t_\perp \approx 0.35$ eV is the interlayer hopping energy, $v_F=(3/2) a t$ denotes the Fermi velocity, with $a=0.142$ nm the lattice spacing and $t \approx 3$ eV the nearest neighbor hopping energy, $p$ is the norm of the momentum vector $\mathbf{p}=(p_x,p_y)$ and $\phi(\mathbf{p})= \arctan \left( p_y/p_x \right)$. Note that if one would have expanded around the $K'$ point, we would have found a Hamiltonian which is the complex conjugate of Eq.~(\ref{ham1}). Since we neglect intervalley interactions, we do not need to take this into account and we simply multiply our results by a factor two.

We perform a change of basis, $ \Psi \to U \Psi$, with
$$ U=\frac{1}{\sqrt{2}}\left( \begin{array}{cccccc} 1 & 0 &0& 0&-1&0 \\ 0 & 1&0&0&0&-1 \\ 1&0&0&0 &1&0 \\ 0 &1& 0 &0&0&1 \\ 0&0&0& \sqrt{2}&0&0 \\ 0&0&\sqrt{2}&0 &0 &0 \end{array} \right), $$
to bring the Hamiltonian into the form $\tilde{H}= \sum \tilde{\Psi}_{
\mathbf{p},\sigma}^\dagger \mathcal{\tilde{H}}(\mathbf{p}) \tilde{\Psi}_{
\mathbf{p},\sigma}$, where
\begin{align}
%\nonumber \tilde{H}&= \sum \tilde{\Psi}^\dagger \mathcal{\tilde{H}} \tilde{\Psi}, \\
\nonumber \tilde{\Psi}_{
\mathbf{p},\sigma}^\dagger &= \frac{1}{\sqrt{2}}( [a_{1,\mathbf{p},\sigma}^\dagger-a_{3,\mathbf{p},\sigma}^\dagger],[b_{1,\mathbf{p},\sigma}^\dagger-
b_{3,\mathbf{p},\sigma}^\dagger],\\
\nonumber &\phantom{=}  [a_{1,\mathbf{p},\sigma}^\dagger+a_{3,\mathbf{p},\sigma}^\dagger], [b_{1,\mathbf{p},\sigma}^\dagger+b_{3,\mathbf{p},\sigma}^\dagger], \\
\nonumber &\phantom{=} \sqrt{2}b_{2,\mathbf{p},\sigma}^\dagger,\sqrt{2}a_{2,\mathbf{p},\sigma}^\dagger), \\
%\nonumber \tilde{\Psi}^\dagger &= ( 1/\sqrt{2}[a_1^\dagger-a_3^\dagger],1/\sqrt{2}[b_1^\dagger-b_3^\dagger],1/\sqrt{2}[a_1^\dagger+a_3^\dagger], \\
%\nonumber &\phantom{=} \qquad 1/\sqrt{2}[b_1^\dagger+b_3^\dagger],b_2^\dagger,a_2^\dagger), \\
\nonumber \mathcal{\tilde{H}}(\mathbf{p}) &=U \mathcal{H}(\mathbf{p}) U^{-1}= \left( \begin{array}{cc} \mathcal{H}_{ml}(\mathbf{p}) &0 \\ 0 & \mathcal{H}_{bl}(\mathbf{p}) \end{array} \right), \\
\nonumber \mathcal{H}_{ml}(\mathbf{p})&= \left( \begin{array}{cc} 0 & v_F p e^{i \phi(\mathbf{p})} \\ v_F p e^{-i \phi(\mathbf{p})} &0 \end{array} \right), \\
\nonumber \mathcal{H}_{bl}(\mathbf{p})&= \\
\nonumber & \left( \begin{array}{cccc} 0 & v_F p e^{i \phi(\mathbf{p})} & -\sqrt{2} t_\perp & 0 \\ v_F p e^{-i \phi(\mathbf{p})} &0&0&0 \\ -\sqrt{2} t_\perp &0 &0 & v_F p e^{i \phi(\mathbf{p})} \\ 0&0& v_F p e^{-i \phi(\mathbf{p})} &0 \end{array} \right).
\end{align}
Thus, the trilayer can be described as a combination of a monolayer and a bilayer with a modified interlayer hopping energy. Note that in the new basis, the basis vectors that are associated with the monolayer part are odd under reflection with respect to the middle plane, while the ones that describe the bilayer are even under this transformation. The hopping parameters $\gamma_2$ and $\gamma_5$ from the Slonczewski-Weiss-McClure (SWM)-model, or a voltage difference between the top and bottom layer break this reflection symmetry and couple the blocks in the trilayer Hamiltonian.\cite{McCa10a} We will neglect those terms here.

Since the Hamiltonian has a block form and we know how to diagonalize the different blocks, it is now a trivial task to bring it into a diagonal form. Using the results from Refs.~\onlinecite{PeCaNe05a} and \onlinecite{NiCaNe06a}, we find that $\mathcal{\tilde{H}}(\mathbf{p})$ can be diagonalized as follows:
\begin{align}
\nonumber \mathcal{D}(\mathbf{p})&= W^\dagger(\mathbf{p})\mathcal{\tilde{H}}(\mathbf{p}) W(\mathbf{p})=W^\dagger(\mathbf{p}) U \mathcal{H}(\mathbf{p}) U^{-1} W(\mathbf{p}) \\
\nonumber &\equiv Z^\dagger(\mathbf{p}) \mathcal{H}(\mathbf{p}) Z(\mathbf{p}), \\
\nonumber W(\mathbf{p}) &= \left( \begin{array}{cc} V(\mathbf{p})& 0 \\ 0& M(\mathbf{p}) \end{array} \right),
\end{align}
where $V(\mathbf{p})$ and $M(\mathbf{p})$ are the matrices that diagonalize the monolayer and bilayer Hamiltonian respectively,
\begin{align}
\nonumber V(\mathbf{p})&=\frac{1}{\sqrt{2}} \left( \begin{array}{cc} -e^{i \phi(\mathbf{p})} & 1 \\ 1 & e^{-i \phi(\mathbf{p})} \end{array} \right), \\
\nonumber M(\mathbf{p})&=M_1(\mathbf{p}) M_2 M_3(\mathbf{p}), \\
\nonumber M_1(\mathbf{p}) &= \left( \begin{array}{cccc} 1 & 0&0&0 \\0& e^{-i \phi(\mathbf{p})} &0&0 \\ 0&0&1&0 \\ 0&0&0& e^{i \phi(\mathbf{p})} \end{array} \right), \\
\nonumber M_2 &= \frac{1}{\sqrt{2}} \left( \begin{array}{cccc} 1 & 0&1&0 \\0& 1 &0&1 \\ 1&0&-1&0 \\ 0&1&0& -1 \end{array} \right), \\
\nonumber M_3(\mathbf{p}) &= \left( \begin{array}{cccc} \cos \varphi(\mathbf{p}) & \sin \varphi(\mathbf{p}) &0&0 \\ -\sin \varphi(\mathbf{p})& \cos \varphi(\mathbf{p}) &0&0 \\ 0&0& \cos \varphi(\mathbf{p})& -\sin \varphi(\mathbf{p}) \\ 0&0& \sin \varphi(\mathbf{p})& \cos \varphi(\mathbf{p}) \end{array} \right).
\end{align}
In the last matrix, $\varphi(\mathbf{p})$ is defined by the relation $\tan[2 \varphi(\mathbf{p})]=v_F \sqrt{2}p/t_\perp$. This result differs by a factor $\sqrt{2}$ from Ref.~\onlinecite{NiCaNe06a} because of the modified interlayer hopping parameter in $\mathcal{H}_{bl}$. The energy bands are given by the nonzero entries of the matrix $\mathcal{D}(\mathbf{p})$,
\begin{align}
\nonumber \mathcal{D}(\mathbf{p}) &=\textrm{diag} \big\{ - v_F p, \, \, v_F p, \, \, [-t_\perp-\xi(p)]/\sqrt{2},  \\
\nonumber &\phantom{=} [-t_\perp+\xi(p)]/\sqrt{2}, \, \,[t_\perp+\xi(p)]/\sqrt{2}, \, \,[t_\perp-\xi(p)]/\sqrt{2} \big\},
\end{align}
where $\xi(p)=\sqrt{t_\perp^2+2 v_F^2 p^2}$.

\begin{figure}
\includegraphics[width=.48\textwidth]{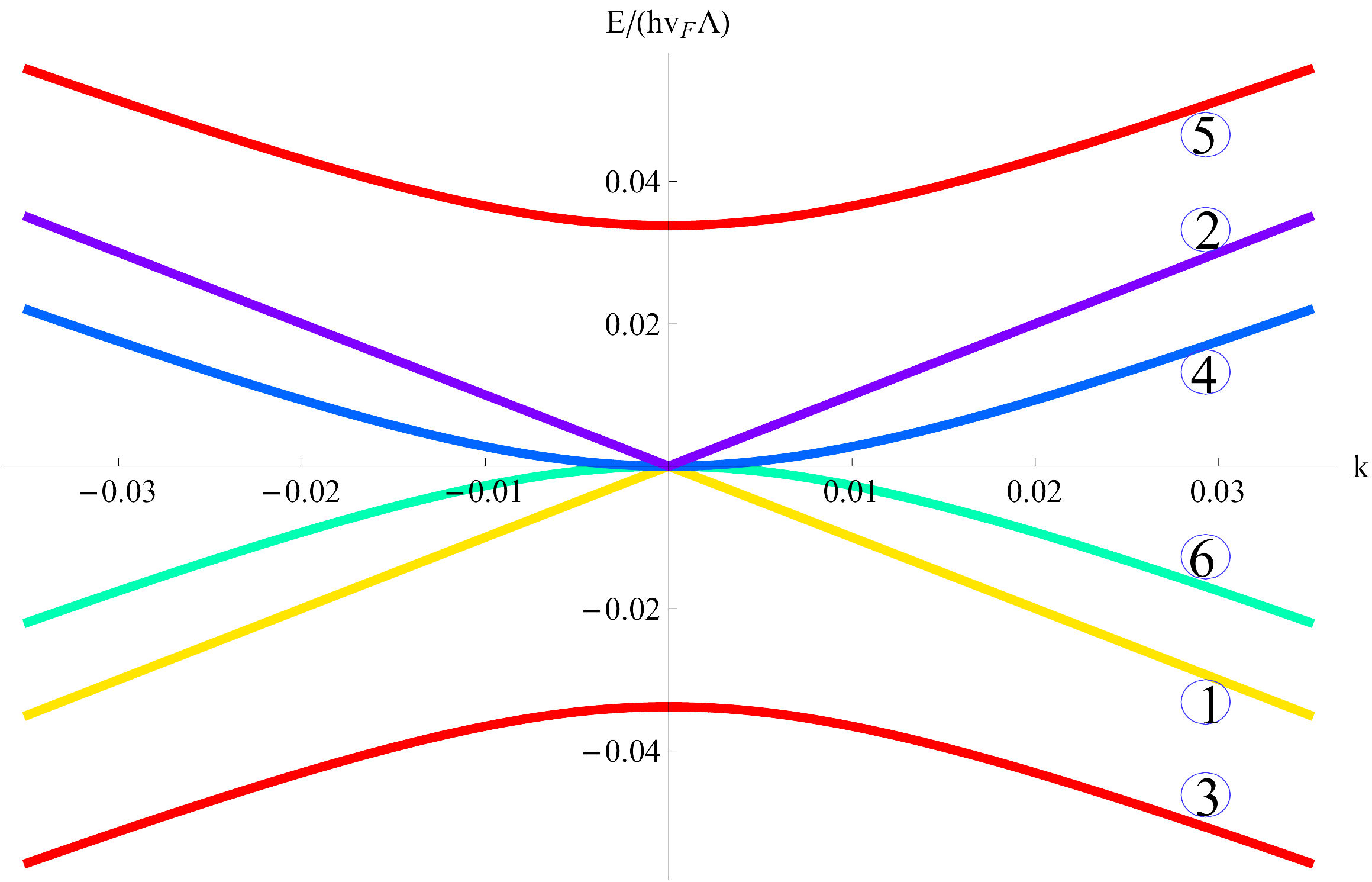}
\caption{ (Color online) The energy spectrum of trilayer graphene. The numbering of the bands is such that  $\Phi_{\mathbf{p},j}^\dagger \Phi_{\mathbf{p},j}=n_j$.}
\label{fig3}
\end{figure}

The next step is to implement the Coulomb interaction in the model. Since we consider only weakly doped trilayers in this paper, the Coulomb interaction is only slightly screened and therefore long ranged,
\begin{align}
\nonumber H_I &= \frac{1}{2} \int d^2 \mathbf{x} \, d^2 \mathbf{y} \big\{ V^\textrm{D}(\mathbf{x}-\mathbf{y}) [ \rho_1(\mathbf{x}) \rho_1(\mathbf{y})+\rho_2(\mathbf{x}) \rho_2(\mathbf{y}) \\
\nonumber &\phantom{=} +\rho_3(\mathbf{x}) \rho_3(\mathbf{y})] +V^{\textrm{ND}}(\mathbf{x}-\mathbf{y})[ \rho_1(\mathbf{x}) \rho_2(\mathbf{y}) \\
\nonumber &\phantom{=}+ \rho_2(\mathbf{x}) \rho_1(\mathbf{y})+ \rho_2(\mathbf{x}) \rho_3(\mathbf{y})+ \rho_3(\mathbf{x}) \rho_2(\mathbf{y})] \\
\label{intrealsp} &\phantom{=}+V^{\textrm{2ND}}(\mathbf{x}-\mathbf{y}) [ \rho_1(\mathbf{x}) \rho_3(\mathbf{y})+\rho_3(\mathbf{x}) \rho_1(\mathbf{y})] \big\},
\end{align}
where $\rho_{i}(\mathbf{x})=\sum_\sigma \left( a_{i,\sigma}^\dagger(\mathbf{x})a_{i,\sigma}(\mathbf{x})+b_{i,\sigma}^\dagger(\mathbf{x})b_{i,\sigma}(\mathbf{x}) \right)$ is the density of electrons in the $i$-th layer and the interaction potentials for the in-plane (D), the nearest-neighbor planes (ND) and the next-nearest-neighbor planes (2ND) are given by
\begin{align}
\nonumber V^{\textrm{D}}(\mathbf{x}-\mathbf{y})&= \frac{e^2}{\epsilon |\mathbf{x}-\mathbf{y}|}, \\
\nonumber V^{\textrm{ND}}(\mathbf{x}-\mathbf{y})&= \frac{e^2}{\epsilon \sqrt{d^2+|\mathbf{x}-\mathbf{y}|^2}}, \\
\nonumber V^{\textrm{2ND}}(\mathbf{x}-\mathbf{y})&= \frac{e^2}{\epsilon \sqrt{4 d^2+|\mathbf{x}-\mathbf{y}|^2}}.
 \end{align}
Here, $e$ is the electron charge, $\epsilon$ the dielectric constant of the substrate (of air in the case of suspended graphene), and $d$ the interlayer distance ($d \approx .32$ nm). The form of $V^\textrm{ND}(\mathbf{x}-\mathbf{y})$ can be understood by recalling that $\mathbf{x}$ is a $2$ dimensional vector. We Fourier transform Eq. (\ref{intrealsp}) and express it in terms of symmetric and anti-symmetric combinations of layer densities,
\begin{align}
\nonumber H_I &= \frac{1}{2A} \sum_\mathbf{q}' \sum_{\alpha= \pm} \big[ \rho_\alpha(\mathbf{q}) V_\alpha(\mathbf{q}) \rho_\alpha (-\mathbf{q}) \\
\label{intterm} &\phantom{=} +\tilde{\rho}_\alpha (\mathbf{q}) V_\alpha (\mathbf{q}) \tilde{\rho}_\alpha (-\mathbf{q})+ \check{\rho}_\alpha (\mathbf{q}) \check{V}_\alpha (\mathbf{q}) \check{\rho}_\alpha (-\mathbf{q})\big],
\end{align}
where the prime on the sum indicates that we omit the $\mathbf{q}=0$ term, since it is canceled by the neutralizing background (Jellium model), $A$ is the area of the unit cell, and the different quantities are defined by
\begin{align}
\label{mid1} \rho_\pm(\mathbf{q})&=\frac{1}{\sqrt{2}} \left[ \rho_1(\mathbf{q})\pm \rho_2(\mathbf{q}) \right], \\
\label{mid2} \tilde{\rho}_\pm(\mathbf{q})&= \frac{1}{\sqrt{2}} \left[ \rho_3(\mathbf{q}) \pm \rho_2 (\mathbf{q}) \right], \\
\label{mid3} \check{\rho}_\pm(\mathbf{q})&= \frac{1}{\sqrt{2}} \left[ \rho_1(\mathbf{q}) \pm \rho_3 (\mathbf{q}) \right], \\
\label{mid4} V_\pm(\mathbf{q})&= \frac{2 \pi e^2}{\epsilon q} \bigg(\frac{1}{2} \pm e^{-q d} \bigg), \\
\label{mid5} \check{V}_\pm(\mathbf{q})&= \frac{2 \pi e^2}{\epsilon q} \bigg(\frac{1}{2} \pm e^{- 2 q d} \bigg).
\end{align}
We want to write this interaction term in the number operators of the energy bands instead of the number operators of the layers. We know how to diagonalize the kinetic term and therefore $\Phi_{\mathbf{p},\sigma} \equiv Z(\mathbf{p})^\dagger \Psi_{\mathbf{p},\sigma}$ are the operators that annihilate particles in the different energy bands. As a result, we obtain, $\left( \Phi_{\mathbf{p},\sigma}^\dagger \Phi_{\mathbf{p},\sigma} \right)_j=n_{j,\sigma}(\mathbf{p})$, the number operator of the $j$-th energy band, where we have to number the bands as in Fig.~\ref{fig3}.
It is convenient to rewrite the density operators in the diagonal basis,
\begin{align}
\label{mid6} \rho_\pm(\mathbf{q})&= \sum_\mathbf{p} \Phi^\dagger_{\mathbf{p}+\mathbf{q}} \chi^\pm(\mathbf{p}+\mathbf{q},\mathbf{p}) \Phi_\mathbf{p}, \\
\label{mid7} \tilde{\rho}_\pm(\mathbf{q})&= \sum_\mathbf{p} \Phi^\dagger_{\mathbf{p}+\mathbf{q}} \tilde{\chi}^\pm(\mathbf{p}+\mathbf{q},\mathbf{p}) \Phi_\mathbf{p}, \\
\label{mid8} \check{\rho}_\pm(\mathbf{q})&= \sum_\mathbf{p} \Phi^\dagger_{\mathbf{p}+\mathbf{q}} \check{\chi}^\pm(\mathbf{p}+\mathbf{q},\mathbf{p}) \Phi_\mathbf{p},
\end{align}
where
\begin{align}
\label{mid9}  \chi^\pm(\mathbf{p}+\mathbf{q},\mathbf{p})&\equiv \frac{1}{\sqrt{2}} Z^\dagger_{\mathbf{p}+\mathbf{q}} \textrm{diag}(1,1,\pm 1,\pm 1,0,0) Z_\mathbf{p}, \\
\label{mid10}  \tilde{\chi}^\pm(\mathbf{p}+\mathbf{q},\mathbf{p})&\equiv \frac{1}{\sqrt{2}} Z^\dagger_{\mathbf{p}+\mathbf{q}} \textrm{diag}(0,0,\pm 1,\pm 1,1,1) Z_\mathbf{p}, \\
\label{mid11}  \check{\chi}^\pm(\mathbf{p}+\mathbf{q},\mathbf{p})&\equiv \frac{1}{\sqrt{2}} Z^\dagger_{\mathbf{p}+\mathbf{q}} \textrm{diag}(1,1,0,0,\pm 1,\pm 1) Z_\mathbf{p}.
\end{align}
Inserting equations (\ref{mid1})-(\ref{mid11}) into the interaction Hamiltonian (\ref{intterm}) yields the interaction term that we use for our calculations. We are only interested in the exchange energy, which is given by
\begin{align}
\nonumber \frac{E_\textrm{ex}}{A} &= -\frac{1}{2} \int \frac{d^2\, \mathbf{p}}{(2 \pi)^2} \frac{d^2\, \mathbf{p}'}{(2 \pi)^2} \sum_{\alpha,i,j,\sigma,a} \\
\nonumber \phantom{+} &\bigg[ \chi^\alpha_{ij}(\mathbf{p}',\mathbf{p}) \chi^\alpha_{ji}(\mathbf{p},\mathbf{p}') V_\alpha(\mathbf{p}'-\mathbf{p}) n_{i,\sigma,a}(\mathbf{p}') n_{j,\sigma,a}(\mathbf{p}) \bigg] \\
\nonumber + &\bigg[ \tilde{\chi}^\alpha_{ij}(\mathbf{p}',\mathbf{p}) \tilde{\chi}^\alpha_{ji}(\mathbf{p},\mathbf{p}') V_\alpha(\mathbf{p}'-\mathbf{p}) n_{i,\sigma,a}(\mathbf{p}') n_{j,\sigma,a}(\mathbf{p}) \bigg] \\
\label{exchangeen} + &\bigg[ \check{\chi}^\alpha_{ij}(\mathbf{p}',\mathbf{p}) \check{\chi}^\alpha_{ji}(\mathbf{p},\mathbf{p}') \check{V}_\alpha(\mathbf{p}'-\mathbf{p}) n_{i,\sigma,a}(\mathbf{p}') n_{j,\sigma,a}(\mathbf{p}) \bigg].
\end{align}
In the sum $\alpha$ takes the values $\pm$; $i$ and $j$ label components, hence run from $1$ to $6$; $\sigma$ sums over spin, and $a$ over the valley index. We neglected the valley index so far since in our case it only gives rise to an extra factor two, as we choose the same pocket structure for both valleys in our studies.

\begin{figure}[t]
\includegraphics[width=.48\textwidth]{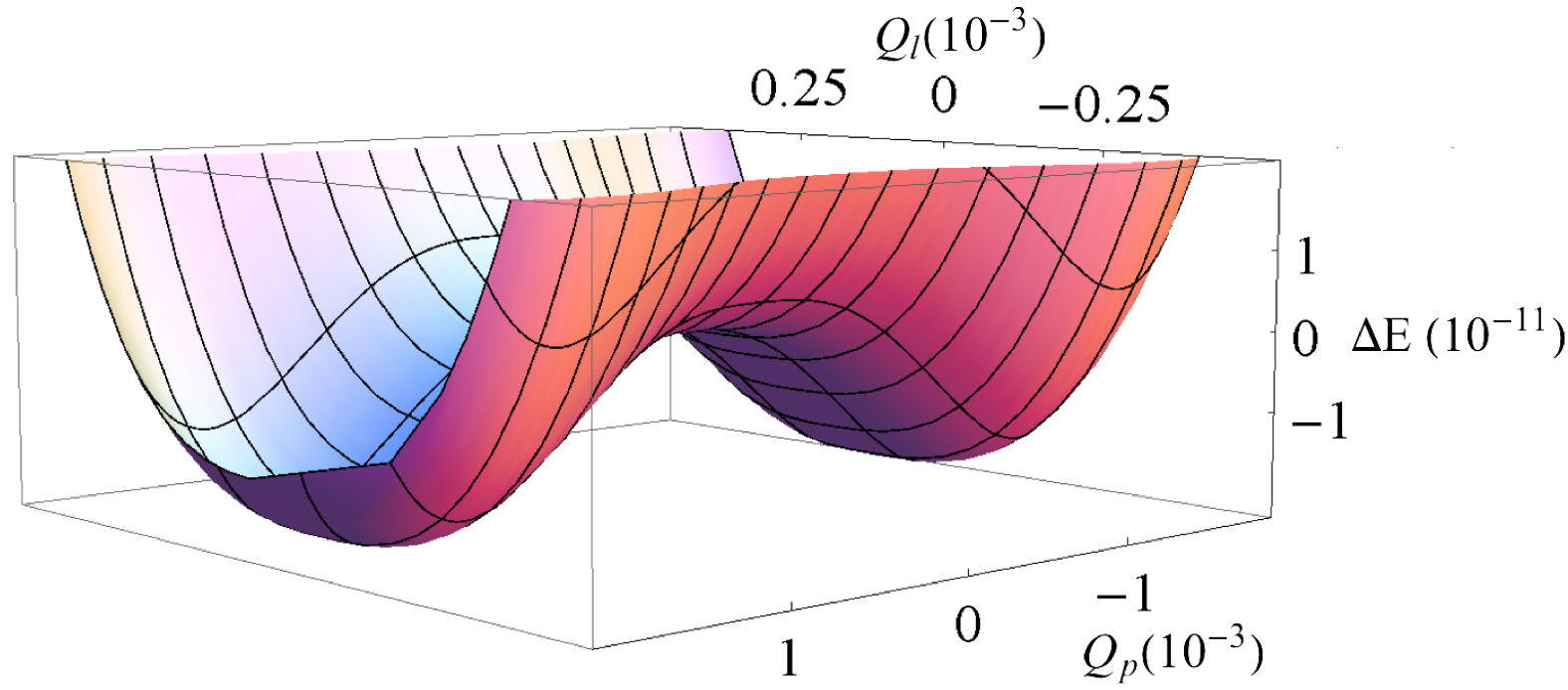}
\caption{(Color online) The energy difference $\Delta E(Q_{lu},Q_{ld},Q_{p u},Q_{p d})$ per unit cell (Eq.~\ref{deltaE4p}) for the undoped trilayer, where we have chosen $Q_{lu}=-Q_{ld} \equiv Q_l$. Because of particle number conservation $Q_{pu}=-Q_{pd}\equiv Q_p$. $\Delta E$ is measured in units of $h v_F \Lambda$, $Q_l$ and $Q_p$ are both measured in units of $\Lambda$. }
\label{fig4}
\end{figure}

\begin{figure}[b]
\includegraphics[width=.48\textwidth]{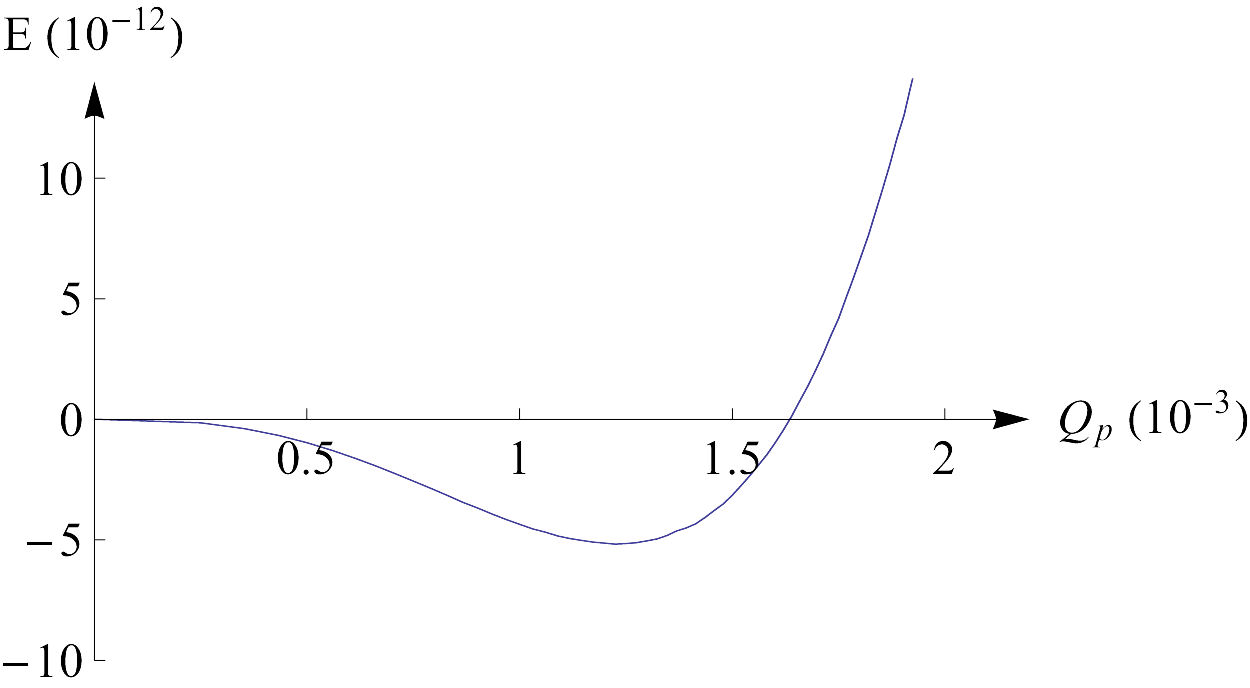}
\caption{(Color online) The energy difference per unit cell $\Delta E(Q_{lu}=0,Q_{ld}=0,Q_{pu}=Q_p,Q_{pd}=-Q_p)$. This is a cross section along the $Q_l=0$ axis of Fig.~\ref{fig4}. $\Delta E$ is measured in units of $h v_F \Lambda$ and $Q_p$ is measured in units of $\Lambda$. }
\label{fig6}
\end{figure}

\section{Ferromagnetic instabilities}
\subsubsection*{Undoped case}

For undoped trilayer graphene, the noninteracting ground state is the configuration in which the three valence bands are completely filled and the conduction bands are completely empty. If an electron or hole pocket forms in one of the bands, this costs kinetic energy. This cost is given by the absolute value of the integral $ \int_0^{E(Q)} dE \, \rho(E) E$, where $Q$ is the pocket size and $\rho(E)$ the density of states. Since for the linear band, $\rho(E) \sim E$ and $E(Q) \sim Q$, one finds that $\Delta E_{\textrm{kin},l} \sim Q^3$, while for the parabolic band, $\rho(E) \sim E^0$, but $E(Q) \sim Q^2$, hence $\Delta E_{\textrm{kin},p} \sim Q^4$. In fact, the changes in kinetic energy for a linear band with pockets of size $Q_l$ and a parabolic band with pockets of size $Q_p$ are
\begin{align}
\nonumber \Delta E_{\textrm{kin},l}(Q_l)&=\frac{A}{6 \pi} \hbar v_F |Q_l|^3, \\
\nonumber \Delta E_{\textrm{kin},p}(Q_p)&=\frac{A} {8 \pi} \frac{(\hbar v_F)^2}{\sqrt{2} t_\perp} |Q_p|^4.
\end{align}
Since $Q_{i}<1$, for $i=l/p$, the creation of linear pockets costs more kinetic energy than the creation of parabolic ones. Trilayer graphene has four energy bands close to the $K$ point, hence there are four different pocket parameters: $Q_{lu}$, $Q_{ld}$, $Q_{pu}$, and $Q_{pd}$, where $l/p$ stands for linear and parabolic bands and $u/d$ for up and down spins. We are assuming long range interactions and are neglecting the short range part, hence there is no intervalley scattering.
%Therefore, the configuration that minimizes the energy at the $K$ point will also be the favorable %configuration at the $K'$ point and the total result only differs by a factor of $2$.
We also assume particle number conservation, thus $Q_{pd}$ is not independent from the other variational parameters. For zero doping one has the constraint:
\begin{align}
\label{construndop} s_{lu} \frac{Q_{lu}^2}{4
\pi}+s_{ld} \frac{Q_{ld}^2}{4 \pi}+s_{pu} \frac{Q_{pu}^2}{4 \pi}+s_{pd} \frac{Q_{pd}^2}{4 \pi}=0,
\end{align}
where $s_{i\sigma}=+1$ for electron-like pockets and $s_{i\sigma}=-1$ for hole-like pockets.

One can now vary the pocket parameters and calculate whether the energy is minimized for nonzero pocket sizes (at zero temperature). Our formalism is build up in such a way that the pocket parameters can take both positive and negative values. A positive $Q$ corresponds to an electron pocket. Hence, the corresponding conduction band (linear/parabolic, up/down) is filled up to momentum $Q$. A negative $Q$ corresponds to hole pockets, i.e. the corresponding valence bands are depleted up to momentum $|Q|$. This method allows us to obtain the exchange integrals for all possible pocket configurations at once.
Using this formalism, we find that the bands fill up according to (see Fig.~\ref{fig3} for the numbering of the bands) ,
\begin{align}
\nonumber n_u(p) &= \left( \begin{array}{c} 1-\Theta(-Q_{lu}-p) \\ \Theta(Q_{lu}-p) \\1 \\ \Theta(Q_{pu}-p) \\0 \\ 1-\Theta(-Q_{pu}-p)) \end{array} \right), \\
\nonumber n_d(p)&= \left( \begin{array}{c} 1-\Theta(-Q_{ld}-p) \\ \Theta(Q_{ld}-p) \\1 \\ \Theta(Q_{pd}-p) \\0 \\ 1-\Theta(-Q_{pd}-p)) \end{array} \right),
\end{align}
where $\Theta$ is the Heaviside step function. Note that one cannot have both electron and hole pockets in the same band at the same time, because if, for example, $Q_{lu}>0$, then $\Theta(-Q_{lu}-p)=0$. Hence, in this case, the linear spin up valence band is completely filled (band 1 in Fig.~\ref{fig3}), while the linear spin up conduction band (band 2 in Fig.~\ref{fig3}) is filled up to momentum $|Q_{lu}|$, corresponding to an electron pocket of size $|Q_{lu}|$.

The integrals that we have to compute have the same structure as the ones in Ref.~\onlinecite{NiCaNe06a}. The expansion in the pocket parameters is highly nontrivial and very lengthy. Since there are three variational parameters, we have performed the integrals numerically.
The expression for the integrals (Eq.~\ref{exchangeen}) has many terms and it is not enlightening to write all of them out.

From this point on, we work in dimensionless units by measuring momenta in units of a cutoff $\Lambda$, which is estimated using a Debye approximation, in which the number of states is conserved in the Brillouin zone: $\Lambda^2= 2 \pi/A$. We measure energies in units of $\hbar v_F \Lambda (A \Lambda^2)=h v_F \Lambda$. This dimensionless energy corresponds with the energy per unit cell in units of $\hbar v_F \Lambda$. Let us also introduce a dimensionless interaction strength $g=e^2/( \epsilon \hbar v_F)$. Furthermore, we set $\Lambda$, $\hbar$ and $t$ equal to unity. %For more details, see the appendix.
Note that the spin-up and spin-down terms decouple. This allows us to calculate
\begin{align}
\nonumber \Delta E(Q_l,Q_p) &=\Delta E_{\textrm{kin}}(Q_l,Q_p)+\Delta E_{\textrm{ex}}(Q_l,Q_p) \\ \nonumber &\equiv \Delta E_{\textrm{kin},l}(Q_l)+\Delta E_{\textrm{kin},p}(Q_p) \\
\nonumber &\phantom{=} +\Delta E_{\textrm{ex},l}(Q_l) +\Delta E_{\textrm{ex},p}(Q_p)  \\
\label{deltaE} &\phantom{=} +\Delta E_{\textrm{ex},\textrm{mixed}}(Q_l,Q_p)\end{align}
on a discrete $N_{l} \times N_{p}$ lattice, where we have chosen the values of the pocket parameters such that their squares lie on an equally spaced grid for reasons which will become clear later. After calculating these data points, one can compute
\begin{align}
\nonumber \Delta E_\textrm{tot}(Q_{lu},Q_{ld},Q_{pu},Q_{pd}) &=\Delta E(Q_{lu},Q_{pu}) \\
\label{deltaE4p} &\phantom{=} +\Delta E(Q_{ld},Q_{pd})
\end{align}
The next step is to select out the points that satisfy the constraint (\ref{construndop}) and find the values of the pocket sizes for which the energy is minimized.

\begin{figure}[t]
\includegraphics[width=.48\textwidth]{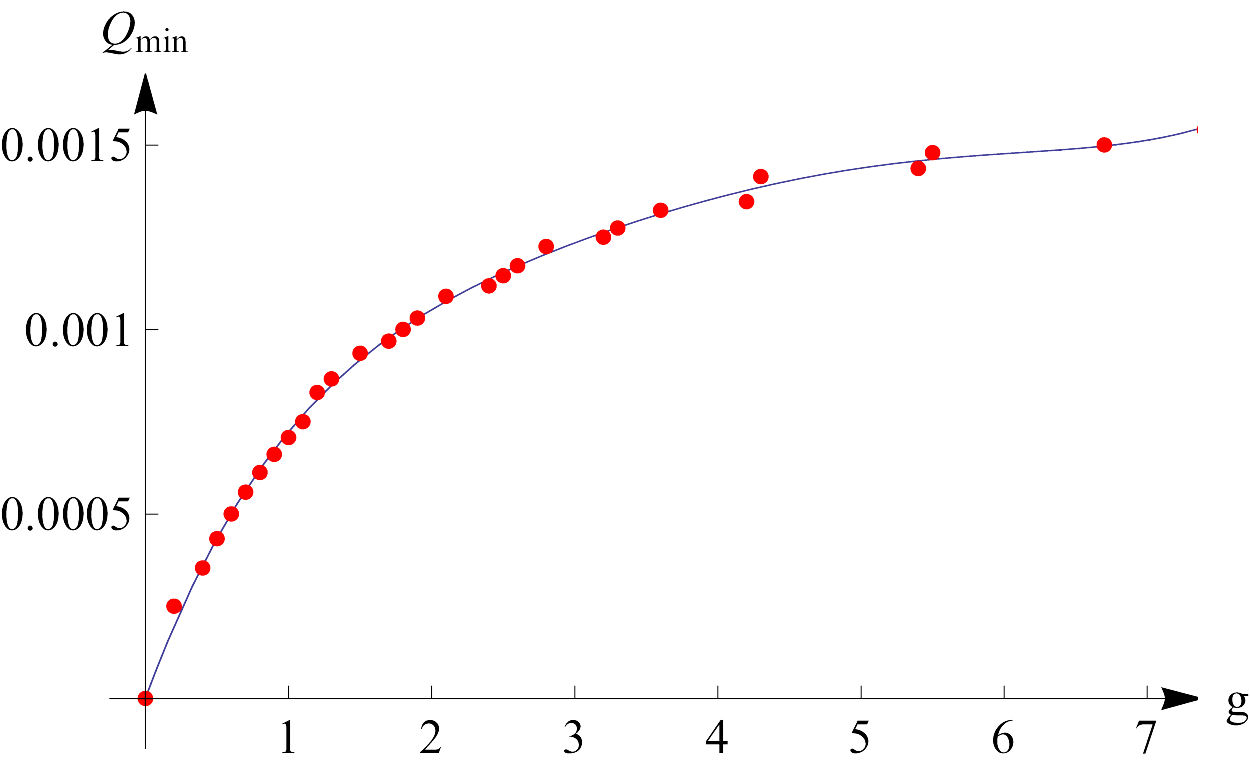}
\caption{(Color online) $Q_{\textrm{min}}$, which is the equilibrium value of $Q_{p}$ as a function of the dimensionless interaction strength $g$. $Q_{\textrm{min}}$ is measured in units of $\Lambda$. The line is a polynomial fit to eighth order in $g$.}
\label{fig5}
\end{figure}

For the undoped case, it turns out that the energy is minimized when the pockets in the linear band are zero, while the pockets in the parabolic band have a nonzero value. This is the result that one obtains if a monolayer and a bilayer are superimposed on each other. There is a priori no reason for this to be the case because in the exchange integrals there appear terms that are mixed in linear and parabolic pocket parameters. However, their contribution is too small to shift the equilibrium value of the pockets in the linear bands away from zero. In Fig.~\ref{fig4}, we have plotted $\Delta E$ as function of $Q_l$ and $Q_{p}$, where $Q_{pu}=-Q_{pd} \equiv Q_p$ due to particle number conservation and we have chosen $ Q_{lu}=-Q_{ld} \equiv Q_l $. Since the spin of the electrons has no preferred direction, one sees two minima in Fig.~\ref{fig4} for $Q_l=0$ and some fixed value of $Q_{p}=\pm Q_{\textrm{min}}$. The energy increases if the linear pocket is chosen to be different from zero, while tuning the parabolic pocket away from zero lowers the energy. Although $\Delta E$ is small (order of 1 meV per square micrometer), the equilibrium sizes of the pockets are significant (see Fig.~\ref{fig6}). The effect is comparable in magnitude with the graphene bilayer. In Fig.~\ref{fig5} we display the equilibrium value for $Q_{p}$ as a function of the interaction strength $g$ (for suspended graphene, $g$ is estimated to be $g \approx 2.3$). The equilibrium value for the linear pocket sizes is zero for this range of the interaction strength.

\subsubsection*{Doped case}

The doped case in trilayer graphene is more subtle than in either a monolayer or a bilayer. For monolayer and bilayers one can dope the system (with electrons or holes) and the bands (spin up and spin down) will fill up to some Fermi energy, corresponding with this particular doping level. This will be the noninteracting ground state for the doped system. In trilayer graphene this is not the case. If one dopes a graphene trilayer such that both the linear and the parabolic band are filled up to some Fermi energy $E_F$, it turns out that due to kinetic energy considerations, this is not a stable state. The kinetic energy is minimized when the parabolic band is filled up differently than the linear band. Alternatively, since for a physical system the Fermi energy has a well defined value, one can interpret this result as a shift of the linear and parabolic energy bands with respect to each other. For our discussion, it is more natural to keep the intersection points of the bands in place and, as a consequence, use different Fermi energies for the parabolic and linear bands. By choosing this interpretation, we allow ourselves to use the formalism developed in the previous section.

\begin{figure}[t]
\includegraphics[width=.48\textwidth]{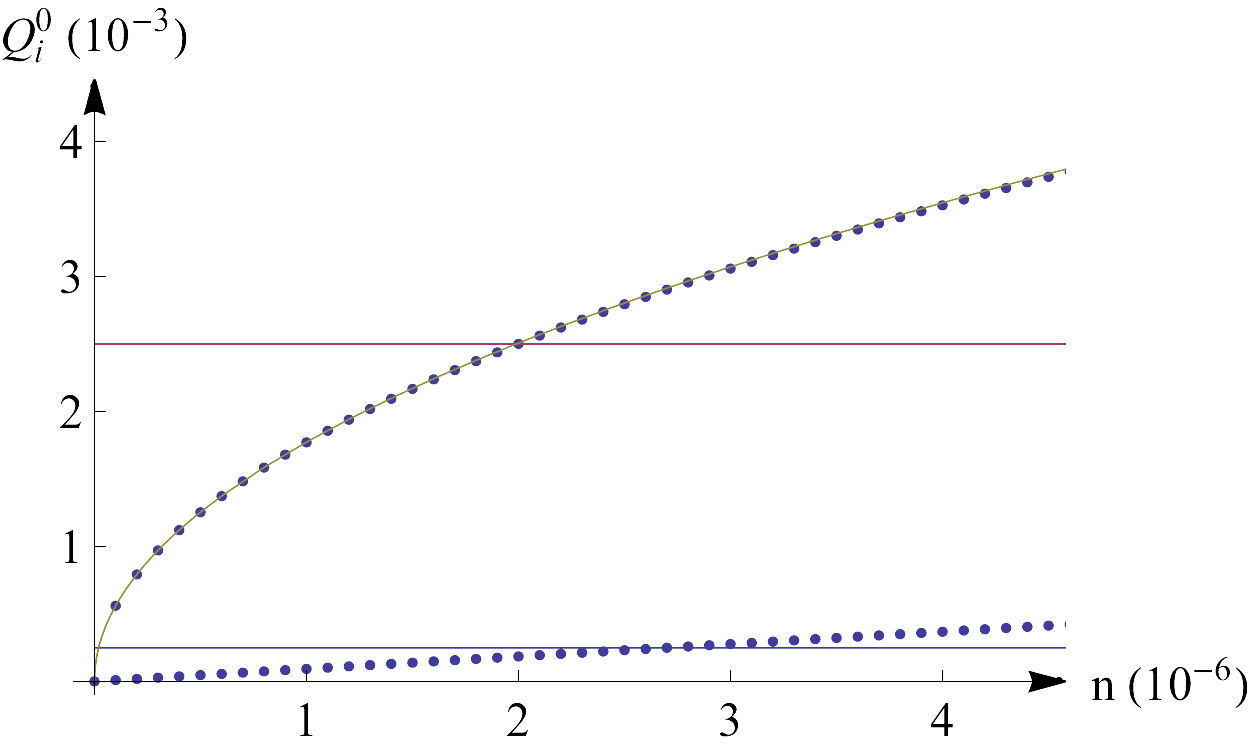}
\caption{(Color online) Plot of $Q_{l}^0$ (blue/black dots) and $Q_p^0$ (grey/grey dots) in units of $\Lambda$ as a function of doping in units of $\Lambda^{-2}$. The red and blue lines mark the interval in which we have chosen our datapoints. The yellow line is a plot of $Q_p^0$ assuming that $Q_l^0=0$.}
\label{fig8}
\end{figure}

Since the kinetic energy cost of filling up the linear band goes as $\approx k^3$, this costs more energy than filling up the parabolic band, for which the energy cost goes as $\approx k^4$ (recall that we work in dimensionless units, such that $k<1$). Let us define $k_F^{l/p,u/d}$ as the momentum to which the linear/parabolic spin up/down band fills up when the kinetic energy is minimized. When there is no interaction present the bands will be spin degenerate. Furthermore, we can use the same formalism as for the undoped case. The difference is that, for $g=0$, the pocketsizes of the bands are equal to $Q_{l/p}^0=k_F^{l/p,u/d}$. Hence, the constraint (\ref{construndop}) now reads,
\begin{align}
 \nonumber &\phantom{=} s_{lu} \frac{Q_{lu}^2}{4
\pi}+s_{ld} \frac{Q_{ld}^2}{4 \pi}+s_{pu} \frac{Q_{pu}^2}{4 \pi}+s_{pd} \frac{Q_{pd}^2}{4 \pi} \\
\label{constrdop} &=s^0_{l} \frac{(Q_{l}^0)^2}{2 \pi}+s^0_{p} \frac{(Q_{p}^0)^2}{2 \pi} \equiv n, \end{align}
where $n$ is the doping level and $s_{l/p}^0$ is the sign of $Q_{l/p}^0$. To determine the values of $Q_{l/p}^0$, one can vary the filling of the bands respecting the constraint and determine for which configuration the kinetic energy is minimized. One can show that $Q_l^0<<Q_p^0$ (see Fig.~\ref{fig8}). In fact, the resolution we use for calculating the integrals is such that $Q_l^0=0$. Note that, although $Q_l^0<<Q_p^0$, the single particle energies associated with these momenta are of the same order of magnitude. The linear band is filled to higher energies than the parabolic one, since the latter is very flat. However, in our discussion this is not relevant because the energies we calculate depend only on momenta and the fact that $Q_l^0=0$ in our formalism barely changes the results. Furthermore, if the effect of interactions on the linear pockets would be such that it would make them larger than the threshold value in Fig.~\ref{fig8}, we would be able to detect it. In the language we proposed in the introduction, this would be a band ferromagnetic state as the bands filled up to different energies, but have no net magnetization.

\begin{figure}[b]
\includegraphics[width=.48\textwidth]{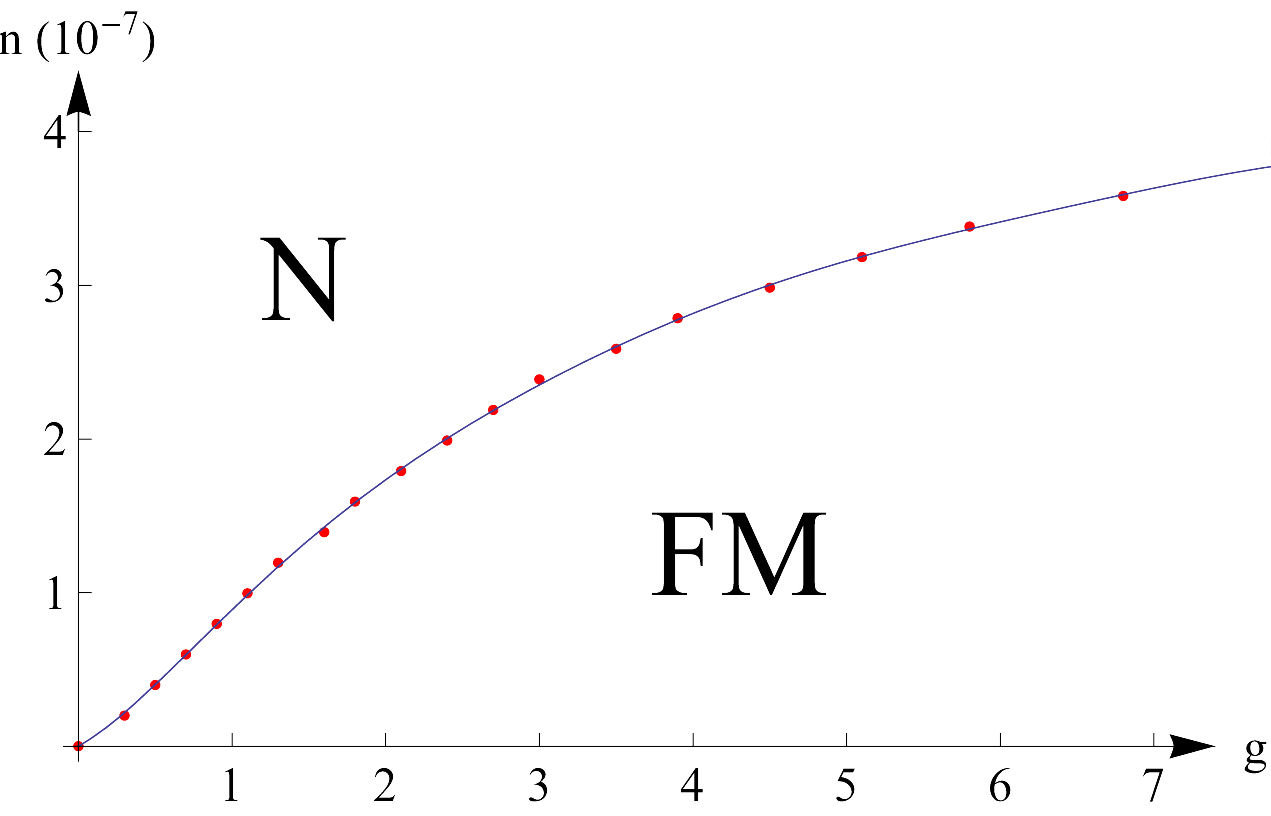}
\caption{(Color online) Phase diagram, doping ($n$) versus interaction strength ($g$). The doping is dimensionless but can be converted to experimental units ($\textrm{cm}^{-2}$) through multiplication with $\Lambda^2$. There is a first order phase transition from the ferromagnetic state (FM) to the normal state (N) as doping is increased.}
\label{fig7}
\end{figure}

In the doped case, the reference state with respect to which we compute energy differences has nonzero kinetic and exchange energies, $E_\textrm{kin}^0=E_\textrm{kin}(Q_l^0,Q_p^0)$ and $E_{\textrm{ex}}^0=2 E_{\textrm{ex}}(Q_l^0,Q_p^0)$. One is now ready to vary the pocket parameters, compute the energies, apply the constraint (\ref{constrdop}), and find the configuration that minimizes the energy
\begin{align}
\nonumber \Delta E &=E_\textrm{kin}(Q_{lu},Q_{pu})+E_\textrm{kin}(Q_{ld},Q_{pd})-E_\textrm{kin}^0 \\
\nonumber &\phantom{=}+E_\textrm{ex}(Q_{lu},Q_{pu})+E_\textrm{ex}(Q_{ld},Q_{pd})-E_\textrm{ex}^0.
\end{align}
The result will depend on the value of the interaction parameter $g$. If the graphene trilayer is doped, the system can still relax into a ferromagnetic state, but a critical interaction strength is needed. This critical value of the interaction increases with doping, as it can be seen in Fig.~\ref{fig7}. The linear bands stay empty (up to our resolution) and the parabolic pockets exhibit a discontinuous jump, indicating a first-order phase transition. This state is both band ferromagnetic, as well as spin ferromagnetic. Note that the jump is such that in one of the parabolic bands hole pockets will occur.

So far we have looked only at configurations in which the pocket sizes are small. Although for the doped case the phase transition is first order, the pocket sizes are small and it is known that in monolayer graphene another first order transition occurs as the interaction strength exceeds some critical value ($g_c \approx 5.3$ for undoped monolayer graphene).\cite{PeCaNe05a} This transition is to a phase in which the monolayer has maximal magnetization. Since, for some purposes, one can regard a trilayer as a combination of monolayer and bilayer graphene, it is natural to look for this transition in a graphene trilayer. Although this transition is theoretically present, we conclude that it can not been seen in any realistic experiment, because the critical coupling is out of any experimental range ($g_c > 200$).

\section{Conclusions}

In this paper, we have determined the ground state of trilayer graphene accounting for the long range Coulomb interaction. We used a formalism in which we could treat electron- and hole-like pockets on the same footing. This allowed us to vary the four pocket parameters (linear/parabolic and spin up/down) to obtain a large dataset. We have chosen the discrete points to lie on a square root profile, so that we had many points that satisfied the constraint (\ref{construndop}) for the undoped system or (\ref{constrdop}) for the doped one.

For the undoped trilayer, we found that the energy is minimized for a configuration in which the linear bands are empty and an electron and a hole pocket occur in the spin up and spin down parabolic bands, which is a spin-ferromagnetic state (Fig.~\ref{fig2}a). Since there is no preferred direction for the spin, this state is doubly degenerate (Fig.~\ref{fig4}). The pockets increase in size when the interaction is tuned to higher values. They are only zero when the interaction vanishes, see Fig.~\ref{fig5}.

The doped trilayer is more subtle, since the noninteracting case is already a band-ferromagnetic state in which the bands (linear/parabolic) fill up differently (Fig.~\ref{fig2}b). We named it a "band-ferromagnetic" state due to the finite polarization in the pseudo-spin degree of freedom associated with parabolic/linear bands. Although in physical systems the bands will shift with respect to each other, resulting in a well defined Fermi energy, we chose to keep the bands fixed and let the bands fill up differently. This gave us two Fermi momenta ($k_F^{l/p}$) and Fermi energies ($E_F^{l/p}$). Although $E_F^p<E_F^l$, the parabolic band is much flatter than the linear one and therefore $k_F^p>>k_F^l$. Our resolution was such that $k_F^l=0$, but this simplification will not affect the results. If the linear pockets exceed the threshold value given by the blue line in Fig.~\ref{fig8} for some value of the interaction strength we would have detected this. It turned out, however, that the linear bands stay empty for all doping levels that we considered. Furthermore, we saw a transition to a spin-ferromagnetic state. In contrast with the undoped case, this state is the ground state only if the coupling exceeds some critical value, which on its turn increases with doping. The doping versus interaction strength phase diagram is shown in Fig.~\ref{fig7}. The phase transition from the normal state (N) to a magnetic state (FM) is first order, i.e. the pocket size jumps discontinuously and the magnetization also exhibits a jump to some nonzero value. Note that this magnetic state is both spin-ferromagnetic and band-ferromagnetic, since the bands fill up to different energies.

We have also looked for a phase transition to a maximally magnetized state, as observed in monolayer graphene. We do not find such a transition for any interaction strength that would be experimentally achievable.

Although the graphene trilayer exhibits some features of both monolayer and bilayer graphene, it is an interesting system on itself and more complex than either of the two. The interplay between the filling of the linear and parabolic bands gives rise to many more possible configurations of the pocket parameters. For example, already in the noninteracting groundstate of the doped trilayer the bands are shifted with respect to each other.

It would be interesting to measure this spectrum in experiments using, for example, angle resolved photo-emission spectroscopy (ARPES). Long range Coulomb interactions can give rise to a ferromagnetic groundstate as it does in bilayer graphene, but will not affect the linear bands. The first order transition as seen in monolayer graphene is not present as a result of interactions between the different bands.

We are aware that next-nearest-neighbor hopping parameters have effects on the energy spectrum that are of comparable magnitude as the effect we describe here.\cite{McCa11} However, if the system is sufficiently doped this will not alter our results. For the undoped case the results may be slightly altered, but our results could definitely be used as a starting point to investigate the full parameter model in more detail.

\section*{Acknowledgements}

The authors acknowledge financial support from the Netherlands Organization for Scientific Research (NWO).

%\bibliographystyle{unsrt}
%\bibliography{bibliography}

\end{document}